\documentclass[% 
superscriptaddress,
 amsmath,amssymb,
 aps,
floatfix,
]{revtex4-2}
\usepackage{graphicx}% Include figure files
\usepackage{dcolumn}% Align table columns on decimal point
\usepackage{bm}% bold math
\usepackage[bordercolor=white,backgroundcolor=gray!30,linecolor=black,colorinlistoftodos]{todonotes}

\usepackage{subfigure}
\usepackage[figuresright]{rotating}
\usepackage{float}
\begin{document}

%\preprint{APS/123-QED}

\title{Small time delay approximation in replicator dynamics}% Force line breaks with \\
\author{Jacek Mi\c{e}kisz}\thanks{miekisz@mimuw.edu.pl}
\affiliation{Institute of Applied Mathematics and Mechanics, University of Warsaw, Banacha 2, 02-097 Warsaw, Poland}
\author{Javad Mohamadichamgavi}\thanks{jmohamadi@mimuw.edu.pl}
\affiliation{Institute of Applied Mathematics and Mechanics, University of Warsaw, Banacha 2, 02-097 Warsaw, Poland}
\author{Raffi Vardanyan}
\affiliation{Institute of Applied Mathematics and Mechanics, University of Warsaw, Banacha 2, 02-097 Warsaw, Poland}

%\date{\today}

\begin{abstract}
We present a microscopic model of replicator dynamics with strategy-dependent time delays. In such a model, new players are born from parents who interacted and received payoffs in the past. 
In the case of small delays, we use Taylor expansion to get ordinary differential equations for frequencies of strategies with time delays as parameters. We apply our technique to get analytic expressions for interior stationary states in two games: Snowdrift and Stag-hunt. We show that interior stationary states depend continuously upon time delays. Our analytic formulas for stationary states approximate well exact numerical results for small time delays.\\[7pt]

\end{abstract}

%\begin{description}
%\item[PACS numbers] 87.23.Kg
%\end{description}

%\pacs{Valid PACS appear here}
\maketitle
%\tableofcontents

% *********************************************************************
%\section{\label{intro}Introduction}
\section{Introduction}
Many biological and socio-economic processes can be modeled by systems of interacting entities such as animals in ecology and evolutionary biology, and people in social systems. It was usually assumed that interactions take place instantaneously and their effects are immediate. In reality, results of biological interactions between individuals may appear in the future, and in social models, individuals or players may act, that is choose appropriate strategies, on the basis of the information concerning events in the past.

It is well known that time delays may cause oscillations in dynamical systems \cite{gyori,Gopalsamy,Kuang,Erneux}. One usually expects that equilibrium of evolving populations, describing coexisting strategies or behaviors, is asymptotically stable for small-time delays and for big ones it becomes unstable.
Here we discuss replicator dynamics of populations of interacting individuals playing two-player games with two strategies and unique interior stationary states.
It describes the time evolution of frequencies of strategies played in the population \cite{Taylor,Hofbauer,Hofbauer2}. Effects of time delays in replicator dynamics were discussed
in \cite{Tao,alboszta,miekisz,Oaku,bodnar2020three,delayspre,Tembine,ijima1,ijima2,moreira}. In \cite{alboszta}, the authors constructed a model, where individuals are born some time after their parents played and received payoffs. There were constructed two coupled equations: one for the frequency of the first strategy and the other one for the size of the population. It was shown that in such a model, oscillations are not possible, the original stationary state is globally asymptotically stable for any time delays.

Here we modify the above model by allowing time delays to depend on strategies played by individuals (strategy-dependent delays were discussed in \cite{Tembine,ijima1,ijima2,moreira}). We consider small delays and use Taylor expansion to get ordinary differential equations for frequencies of strategies with time delays as parameters. We apply our technique to get analytic expressions for interior stationary states in two games: Snowdrift and Stag-hunt. We show that interior stationary states depend continuously upon time delays. Our analytic formulas for stationary states approximate well exact numerical results obtained in \cite{delayspre} for small time delays.

In Section 2, we construct replicator dynamics with strategy-dependent time delays. Section 3 contains a derivation of replicator dynamics for small delays. In Section 4, we apply our technique in two examples: Snowdrift and Stag-hunt games. Discussion follows in Section 5.

\section{Replicator dynamics with strategy-dependent time delays}
\label{sec:2}

We consider games with two pure strategies; denote them by C and D. In discrete moments of time, individuals compete in pairwise
contests and the outcome is given by the following
payoff matrix:

\begin{center}
\begin{tabular}{ c c c c }
 &      & C & D \\ 
    & C & a & b \\
U = & & &     \\  
    & D & c & d    
\end{tabular}
\end{center}

where the $i, j$ entry, $i, j = C, D$, is the payoff of the first (row) player when he plays the strategy $i$ and the second
(column) player plays the strategy $ j$. We assume that both players are the same and hence payoffs of the
column player are given by the matrix transposed to U; such games are called symmetric. From now on we will
assume that  $a<c$ and $d<b$ or $c<a$ and $b<d$ so there is a unique mixed Nash equilibrium,  $\bar x = \frac{{ b-d }}{{b - d + c - a}}$,
an interior stationary state in the replicator dynamics, respectively asymptotically stable or unstable one. We assume that each player interacts with all other ones 
and receive an average payoff with respect to the structure of the population, i.e. the proportion of the population playing each strategy. 
We interpret payoffs as a number of offspring that an individual has after a contest, the offspring inherits the strategy of an ancestor. 

We assume that during a very small-time interval of length $\varepsilon$, only an $\varepsilon$-fraction of the population can manage to pair with partners and play the game. We assume that players do not get payoffs immediately - new players are born  $\tau$ units of time after parents interacted and received payoffs. We also assume that time delays are strategy dependent, we denote them by $\tau_{C}$ and $\tau_{D}$.

Let $p_{i}(t)$, $i=C, D,$ be the number of individuals who play strategies $C$ and $D$ respectively at the time $t$. Then the total number of players will be
$p(t)=p_{C}(t)+p_{D}(t)$ and the fraction of the population playing strategy $C$ will be $x(t)=\frac{p_{C}(t)}{p(t)}$.
Average payoffs which players get when playing the strategies $C$ and $D$, are given by $U_{C}(t)= ax(t)+ b(1-x(t))$ and  $U_{D}(t)= cx(t)+d(1-x(t))$ respectively. 

With all these notations and assumptions in mind, we propose the following equations to describe our model:

\begin{equation}
       p_{i}(t + \varepsilon) = (1 - \varepsilon) p_{i}(t) + \varepsilon p_{i}(t-\tau_{i})U_{i}(t-\tau_{i}), \label{bdiscrete1}
\end{equation}

$i~= C, D$. Then for the size of the whole population we get 

\begin{equation}
      p(t + \varepsilon)= (1 - \varepsilon) p(t) + \varepsilon \Bigl(p_{C}(t-\tau_{C})U_{C}(t-\tau_{C})+ p_{D}(t-\tau_{D})U_{D}(t-\tau_{D})\Bigr) .\label{bdiscrete2}   
\end{equation}

We divide (\ref{bdiscrete1}) by (\ref{bdiscrete2}) for i = C, obtain the equation for $x(t +\varepsilon)$, subtract $x(t)$, divide the difference by $\varepsilon$, take
the limit $\varepsilon \rightarrow 0$, and get an equation for the frequency of the first strategy,

\begin{equation}
    \frac{dx}{dt}= \frac{p_{C}(t-\tau_{C})U_{C}(t-\tau_{C})(1-x(t)) - p_{D}(t-\tau_{D})U_{D}(t-\tau_{D})x(t)}{p(t)}.\label{bxeq}
\end{equation}

Let us notice that unlike in the standard replicator dynamics, the above equation for the frequency of the first strategy
is not closed, one needs equations for populations sizes. From (\ref{bdiscrete1}) and (\ref{bdiscrete2}) we get

\begin{equation}
  \frac{d p_{i}(t)}{dt}= - p_{i}(t) +  p_{i}(t - \tau_{i})U_{i}(t-\tau_{i}), i~= C, D \label{bpieq}   
\end{equation}

\begin{equation}
 \frac{dp(t)}{dt}= - p(t) +  p_{C}(t - \tau_{C})U_{C}(t-\tau_{C})+ p_{D}(t - \tau_{D})U_{D}(t-\tau_{D}). \label{bpeq}   
\end{equation}

In \cite{delayspre}, a transcendental equation for stationary states of (\ref{bxeq}) was obtained and then it was solved numerically for various games. Here we derive an approximate replicator equation 
for small time delays which enables us to get an analytical expression for the interior stationary state.

\section{Small time delays approximation}
\label{sec:3}

We begin by presenting (\ref{bxeq}) in a different form. We insert (\ref{bpieq}) into (\ref{bxeq}) and after some simplifications we get the following equation:

\begin{equation}
\frac{dx}{dt} = \frac{1}{p(t)} [\frac{dp_{C}(t)}{dt} (1-x(t)) - \frac{dp_{D}(t)}{dt} x(t)]. \label{sdxeq}
\end{equation}

Now we Taylor expand the right part of (\ref{bpieq}), keep first powers of $\tau_{C}$ and $\tau_{D}$, and get the following equation:

\begin{equation}
   \frac{dp_{i}(t)}{dt} = - p_{i}(t) + p_{i}(t) U_{i}(t) -
 \tau_{i}[\frac{dp_{i}(t)}{dt} U_{i}(t) + p_{i}(t)\frac{dU_{i}(t)}{dt}],\label{sdpieq} 
\end{equation}

and hence

\begin{equation}
       \frac{dp_{C}(t)}{dt} = - p_{C}(t) + (a-b) x(t) p_{C}(t)+ b p_{C}(t) -
 \tau_{C}[\frac{dp_{C}(t)}{dt}((a-b)x(t)+b) +(a-b) p_{C}(t)\frac{dx(t)}{dt}],\label{sdpAeq}
 \end{equation}
 \begin{equation}
\frac{dp_{D}(t)}{dt} = - p_{D}(t) + (c-d) x(t) p_{D}(t)+ d p_{D}(t) -
 \tau_{D}[\frac{dp_{D}(t)}{dt}((c-d)x(t)+d) +(c-d) p_{D}(t)\frac{dx(t)}{dt}].\label{sdpBeq}
\end{equation}

We solve (\ref{sdpAeq}) and (\ref{sdpBeq}) for derivatives of $p_{C}(t)$ and $p_{D}(t)$,
\begin{equation}
    \frac{dp_{C}(t)}{dt} = px \frac{-1+b + (a-b)x - (a-b)\tau_{C}\frac{dx}{dt} }{1+ \tau_{C}(a-b)x + b\tau_{C}},\hspace{30pt}\label{sdpAeq1}
\end{equation}
 \begin{equation}
      \frac{dp_{D}(t)}{dt} = p(1-x)  \frac{-1+d + (c-d)x - (c-d)\tau_{D}\frac{dx}{dt} }{1+ \tau_{D}(c-d)x + d\tau_{D}}.\label{sdpBeq1}
 \end{equation}

Now we insert (\ref{sdpAeq1}) and (\ref{sdpBeq1}) into (\ref{sdxeq}), solve it for $\frac{dx}{dt}$ and get one ordinary differential equation for the frequency of the first strategy with $\tau_{C} $ and $\tau_{D} $ as parameters,

\begin{equation}
    \frac{dx}{dt} =
 \frac{x(1-x)}{K} \left[ 
\frac{-1+b + (a-b)x}{1+ \tau_{C}(a-b)x + b\tau_{C}} - 
 \frac{-1+d + (c-d)x}{1+ \tau_{D}(c-d)x + d\tau_{D}}
\right],\label{sdxSol}
\end{equation}
where
\[
   K = 1- x(1-x) \left[ 
\frac{-(a-b)\tau_{C}}{1+ \tau_{C}(a-b)x + b\tau_{C}} + 
 \frac{(c-d)\tau_{D}}{1+ \tau_{D}(c-d)x + d\tau_{D}}
\right].  
\]

Now it is easy to get an equation for the stationary state,

\begin{equation}\label{replicator}
\alpha x^2 + \beta x + \gamma = 0,
\end{equation}

where

\begin{equation*}
\begin{cases}
    \alpha = (\tau_{D} - \tau_{C})(a-b)(c-d),\\
        \beta = \tau_{D}[(a-b)d + (b-1)(c-d)]+\tau_{C}[(d-c)b -(d-1)(a-b)] +a-b-c+d, \\
        \gamma = \tau_{D}d(b-1) + \tau_{C}b(1-d) + b-d
  \end{cases}
\end{equation*}

We solve (\ref{replicator}) on the interval $[0,1]$ and get a formula for the stationary state,

\begin{equation}\label{sdxSSSol}
\bar x = \frac{ - \beta \pm \sqrt{\beta^2 -4\alpha \gamma}}{2\alpha}.
\end{equation}

We see that frequencies of strategies in the stationary state depend continuously on time delays, $\tau_{C}$ and $\tau_{D}$, not only on their difference.

It is easy to see that when $\tau_{C}$ is equal to $\tau_{D}$, then there is only one solution of (\ref{replicator}), namely  $\bar x = \frac{{ b-d }}{{b - d + c - a}}$ 
as in the replicator dynamics without time delays.

\section{Examples}
\label{sec:4}

Here we will analyze dependence of stationary states on time delays for Snowdrift and Stag-hunt game. 
We will also compare our approximate analytical results with numerical solutions for stationary states obtained in \cite{delayspre}.

\subsection{Snowdrift game}

Snowdrift game describes interactions between two car drives caught in a snow blizzard. They can cooperate and  clear together the road from snow. However one of the drivers can defect  and wait until the second one will do the job. However if both of them defect they will never go home. We see that it is profitable for a driver to choose the strategy different from that of the other one. That leads to a replicator dynamics with a stable interior state - a stable coexistence of both strategies.

Snowdrift game is usually parameterized as follows: 
\begin{center}
\begin{tabular}{ c c c c }
 &      & C & D \\ 
    & C & b-c/2 & b-c \\
$U_1$ = & & &     \\  
    & D& b & 0    
\end{tabular}
\end{center}

In this matrix strategy $C$ stands for cooperation and $D$ for defection, $b$ stands for the benefit and $c$ is the cost that a cooperator pays, we of course assume that $b > c$. 

For such a game we have:

\begin{equation*}
\begin{cases}
    \alpha = (\tau_{D} - \tau_{C})bc/2,\\
        \beta = \tau_{D}b(b-c-1) - \tau_{C}(b(b-c) - c/2) + c/2 - b,\\
        \gamma = \tau_{C}(b - c) + b - c
  \end{cases}
\end{equation*}

and for $b = 6$ and $c = 4$

\begin{equation}\label{sdxSSSolHD}
\begin{split}
\bar x = \frac{-(6\tau_{D} - 10\tau_{C} - 4)-\sqrt{(6\tau_{D} - 10\tau_{C} - 4)^2 - 48(\tau_{D} - \tau_{C})(2\tau_{C} + 2)}}{24(\tau_{D} - \tau_{C})}
\end{split}
\end{equation}

\begin{figure}
   \centering
      \includegraphics[width=.6\linewidth]{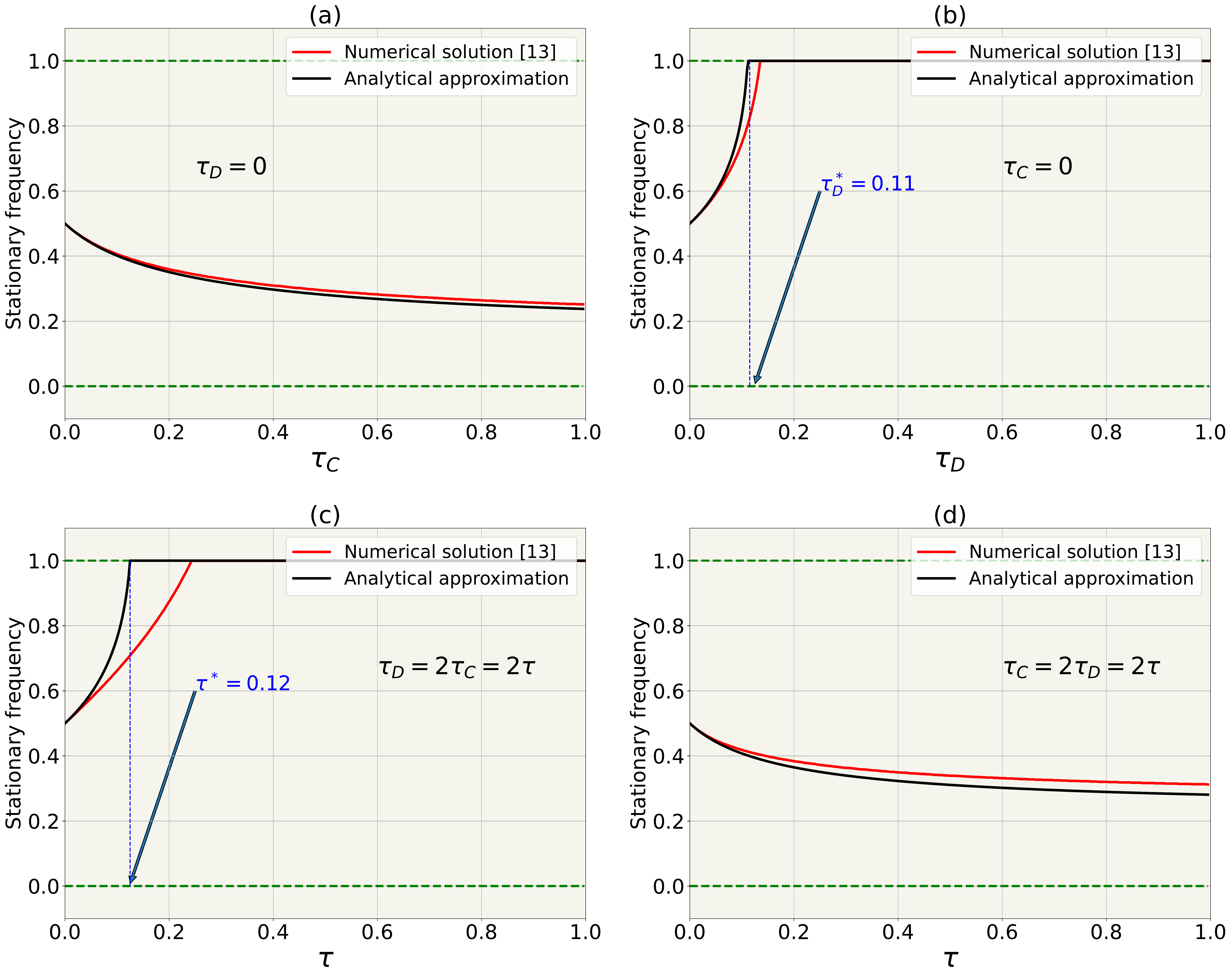}
      \label{fig1}
   \caption{Dependence of the stable interior stationary state on time delays in the Snowdrift game with $b=6, c=4$; we present the small-time approximation $\bar x$ and the numerical solution of the equation for stationary state \cite{delayspre}.  (a) stationary state as a function of $\tau_C$, when $\tau_D =0$, (b) stationary state as a function of $\tau_D$, when $\tau_C =0$, (c) stationary state as a function of $\tau$, when $\tau_D=2\tau$ and $\tau_C=\tau$, (d) stationary state as a function of $\tau$ when $\tau_C=2\tau$ and $\tau_D=\tau$.}
\end{figure}

In Fig. 1 we show how the stationary state, that is the frequency of the cooperation, depends on time delays. We compare our small-delay approximation with numerical results obtained in \cite{delayspre}, the agreement is quite good. We see that the bigger is a time delay of a given strategy, smaller is its proportion in the population. Let us note that delay in the strategy $D$ affects the stationary state much more dramatically than delay in the strategy $C$. We see that for $\tau_C=0$, the stable interior stationary state increases as $\tau_D$ 
increases and it disappears as $\tau_D>\tau_D^* = 1/9$. In general, it follows from (\ref{sdxSSSolHD}) that a population consisting of just of C-players
 becomes globally asymptotically stable 
for $\tau_D \geq \tau_D^*=(10\tau_C+1)/9$.

\subsection{Stag-hunt game}

Stag-hunt game describes the competition between two hunters: if they coordinate their actions, they get a stag, but if one of them defects, he will get a hare and the other one will stay with nothing. It follows that it is profitable for hunters to choose the same strategy. The game belongs to a class of games with two stable Nash equilibria and an unstable interior point (a mixed Nash equilibrium) in the replicator dynamics. Here we will consider the following payoff matrix:
\begin{center}
\begin{tabular}{ c c c c }
 &      & C & D \\ 
    & C & s & 0 \\
$U_2$ = & & &     \\  
    & D & h & h    
\end{tabular}
\end{center}
\begin{figure}
   \centering
      \includegraphics[width=.6\linewidth]{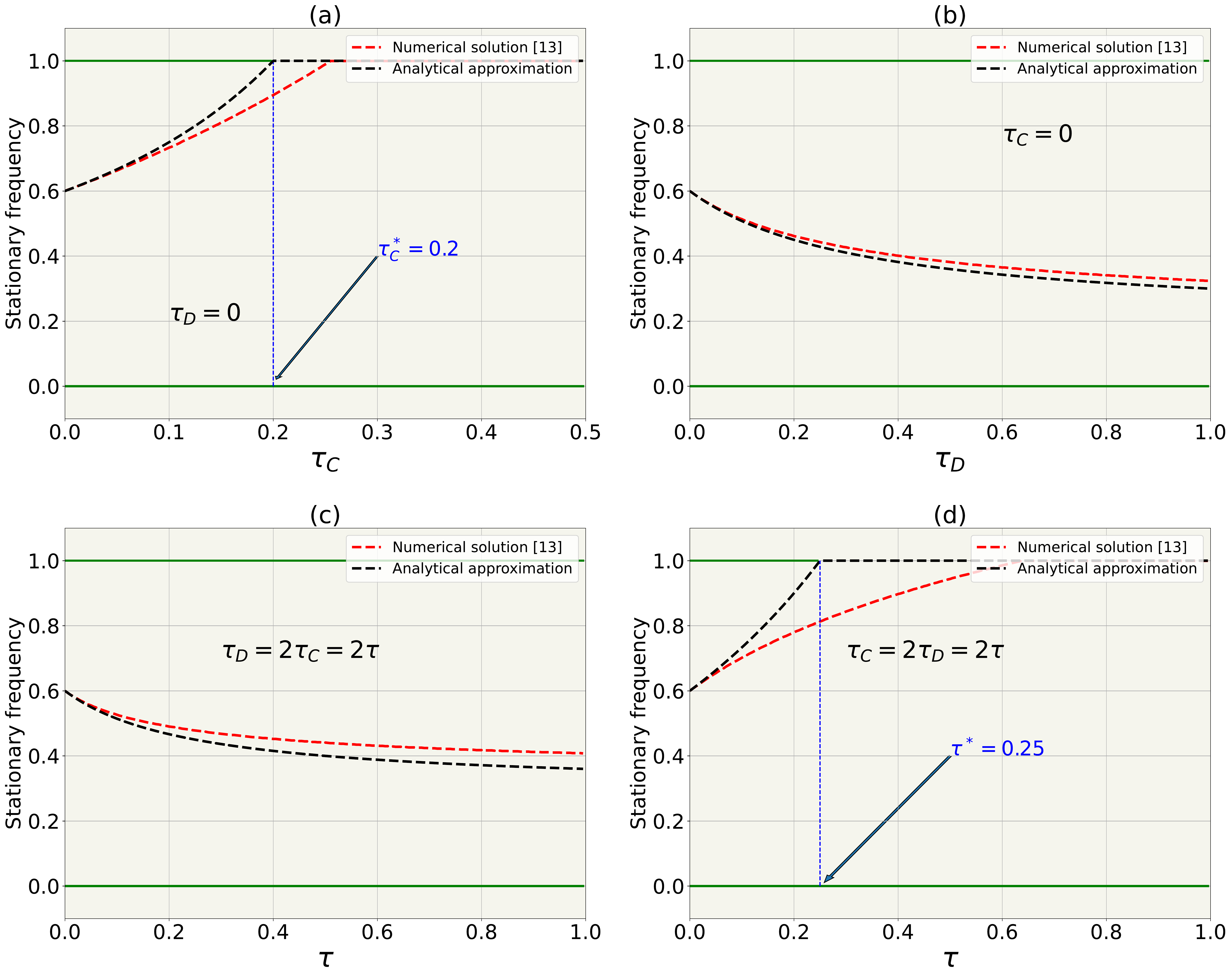}
      \label{fig3}
   \caption{Dependence of the unstable interior stationary state on time delays in the Stag-hunt game with $s=5, h=3$; we present the small-time approximation $\bar x$ and the numerical solution from \cite{delayspre}. (a) stationary state as a function of $\tau_C$, when $\tau_D = 0$, (b) stationary state as a function of $\tau_D$, when $\tau_C =0$, (c) stationary state as a function of $\tau$, when $\tau_D=2\tau$ and $\tau_C=\tau$, and (d) stationary state as a function of $\tau$ when $\tau_C=2\tau$ and $\tau_D=\tau$.}
\end{figure}
Now we get

\begin{equation*}
\begin{cases}
        \alpha = \alpha = 0,\\
        \beta = \beta = \tau_{D}sh  - \tau_{C}s(h-1)+s,\\
        \gamma = \gamma = -\tau_{D}h -h.
  \end{cases}
\end{equation*}
  
Thus the stationary state is given by

\begin{equation}\label{sdxSSSolSH}
\bar x = \frac{h(\tau_{D} + 1)}{s + hs(\tau_{D} - \tau_{C}) + s\tau_{C}}.
\end{equation}

Now we choose $s = 5$ and $h = 3$. In Fig. 2, we show the dependence of the stationary state on time delays. Again we see that our small-delay approximation is quite good. Now the bigger is a time delay of a given strategy, smaller is its basin of attraction. We see that for $\tau_C=0$, the stable interior stationary state increases as $\tau_D$ increases and it disappears as $\tau_C>\tau_C^* = 0.2$. In general, it follows from (\ref{sdxSSSolSH}) that a population consisting of just D-players becomes globally asymptotically stable for $\tau_C \geq \tau_C^*=\frac{12\tau_D+2}{10}$.
Delay of the strategy $C$ affects the stationary state more significantly than delays in the strategy $D$.

\section{Discussion}
\label{sec:5}

Replicator dynamics with strategy-dependent time delays was studied in \cite{delayspre}. The authors derived a transcendental equation for a stationary state and solved it numerically for various games. Here we introduced a small-delay approximation for time-delayed differential equations. This enabled us to approximate the system of replicator equations for the strategy frequency and the population size by an ordinary differential equation for the strategy frequency with time delays as parameters. In this way we obtained an analytic expression for the stationary state. We applied our technique to the Snowdrift and Stag-hunt game. Our analytic formulas approximate well exact numerical solutions for small time delays.

We showed that stationary frequencies of strategies depend continuously on time delays. Moreover, for the Snowdrift game, the frequency of a strategy in the stationary state is a decreasing function of its delay; for the Stag-hunt game, a basin of attraction of a strategy is a decreasing function of its delay. Such results were of course already presented in \cite{delayspre}.

It would be interesting to analyze time-delayed replicator dynamics taking into account stochastic effects resulting from mutations and a random character of interactions. In the latter case, individuals play with particular opponents and not against the average strategy like in the standard replicator dynamics. We would also like to study time-delay effects in evolutionary games in finite populations. Some results were already presented in \cite{delaywalk}. 
\vspace{3mm}

{\bf Acknowledgments}: This project has received funding from the European Union’s Horizon 2020 research and innovation programme under the Marie Sk\l odowska-Curie grant agreement No 955708. J. Mi\c{e}kisz and R. Vardanyan thank the National Science Centre, Poland, for a financial support under Grant No. 2015/17/B/ST1/00693.

\end{document}